\documentclass[submission,copyright,creativecommons]{eptcs}
\providecommand{\event}{FROM 2023} % Name of the event you are submitting to

\usepackage{iftex,xcolor,fancyvrb}
\ifpdf
  \usepackage{underscore}         % Only needed if you use pdflatex.
  \usepackage[T1]{fontenc}        % Recommended with pdflatex
\else
  \usepackage{breakurl}           % Not needed if you use pdflatex only.
\fi
\usepackage{amsmath,amssymb,stmaryrd,txfonts}

\newcommand{\N}{\mathbb{N}}
\newcommand{\lub}{\mathit{lub}}

\newtheorem{definition}{Definition}

\newtheorem{theorem}{Theorem}
\newtheorem{example}{Example}
\newtheorem{corollary}{Corollary}

\title{While Loops in Coq}
\author{%
David Nowak, CNRS\footnote{Univ. Lille, CNRS, Centrale Lille, UMR 9189 CRIStAL, F-59000 Lille, France}
\and
Vlad Rusu, Inria\footnote{Univ. Lille, Inria, Centrale Lille, UMR 9189 CRIStAL, F-59000 Lille, France}
}
\def\titlerunning{While Loops in Coq}
\def\authorrunning{D. Nowak \& V. Rusu}

\begin{document}
\maketitle

\begin{abstract}
While loops are present in virtually all imperative
programming languages. They are important both for practical reasons
(performing a number of iterations not known in advance) and
theoretical reasons (achieving Turing completeness).
 In this paper we propose an approach for
incorporating while loops in an imperative language shallowly embedded in
the Coq proof assistant. The main difficulty is that proving
the termination of while loops is nontrivial, or impossible
in the case of non-termination, whereas Coq only accepts
 programs endowed with termination proofs. Our solution is based on
 a new, general method for defining possibly non-terminating recursive functions in Coq.
We illustrate the
approach by 
proving termination and partial correctness of a program on linked
lists. \end{abstract}

\section{Introduction}

The definition of recursive functions in the Coq proof assistant~\cite{BC04} is subject to certain restrictions to ensure their termination, which is essential for the  consistency of Coq's underlying logic. Specifically, recursive calls must be made on strict subterms, effectively ensuring that the computation eventually reaches a base case. Alternatively, users have the option to prove that a specific quantity strictly decreases according to a well-founded order. In such cases, Coq can automatically transform the recursive calls into strict subterm calls, using a so-called accessibility proof to guarantee termination. Adhering to these constraints eliminates the risk of infinitely many calls, thereby ensuring that  functions terminate.

An alternative, somewhat ad-hoc strategy is to introduce an additional natural-number argument called the~\textit{fuel}. The fuel's value is decremented with each recursive call, thereby guaranteeing finitely many recursive calls, hence, termination. However, a crucial concern arises as one must  supply enough fuel so that termination does not disrupt the intended computation of the program by occurring too early.

In this paper we present a novel approach to defining possibly partial recursive functions in Coq while achieving separation of concerns: write the program first, and prove its properties (including termination) later.
In broad terms our technique consists in  providing an infinite amount of fuel for recursive functions. By doing so the function can proceed with its computations without risk of exhausting its fuel.

As a result, this approach empowers developers to focus on the core logic of the recursive function, separate from the termination concern, streamlining the development process and enhancing the modularity and readability of the code. 

A key property  guaranteed by our technique
 is that, given the \emph{functional}  of  the recursive function under definition (i.e., an abstract description of the function's body), the resulting function is the \emph{least fixpoint} of its functional. We prove this  general result under mild constraints on the functional -  it must be monotonic and, in some sense described precisely in the paper, must preserve continuity.

 The method is applied to while loops in an imperative shallowly embedded in Coq. By proving
that the functional of  while loops is monotonic and continuity-preserving we obtain while loops as least fixpoints of their functionals.
 This enables programmers to
construct imperative programs featuring  arbitrary while loops and provides them with tools for reasoning about  loops.
Specifically, the least-fixpoint property  is used for proving that a while loop terminates if and only if there exists some finite amount of fuel for which it returns the desired result; hence, a termination proof can proceed by induction on the  fuel value, once an adequate instantiation for this  known-to-exists value is chosen.

Subsequently, we proceed to establish a Hoare logic system, which serves as a formal tool for proving the partial correctness of programs. 
In essence the Hoare logic provides a systematic and rigorous approach to program verification, where one defines preconditions and postconditions that govern the state of the program before and after its execution. Through these assertions one can verify that the program's execution leads to the desired outcomes, establishing its partial correctness. Here, again, the property of being a fixpoint 
is used for proving the soundness of the while loop's \emph{Hoare triple}.

Finally, the least-fixpoint property of the partial functions being defined ensures that the functions,  defined abstractly using order theory, are, as mathematical functions,  the same as the ones that Coq would have generated, had it not been constrained by its logic into rejecting the functions as  partial.

\paragraph{Outline}

In Section~\ref{sec:monad} we introduce the reader, termination, and state monads, which serve the purpose of writing imperative programs in the Coq proof assistant.
Moving forward to Section~\ref{sec:recursion} we present our method for defining possibly nonterminating recursive functions and demonstrate its application to the definition of while loops.
In Section~\ref{sec:correctness} we define a monadic Hoare logic and illustrate its effectiveness by applying it to a program that computes the length of a linked list.
In Section~\ref{sec:termination} we address the issue of proving  termination.
We compare with related work in Section~\ref{sec:rel} and conclude in Section~\ref{sec:conclusion}.

The Coq development corresponding to this paper is available at~\url{https://tinyurl.com/2p93uwdj}.

\section{Monads for Possibly Nonterminating Stateful Computation}
\label{sec:monad}

We consider a subset of Gallina (the programming language of Coq) 
 expressive enough for shallowly embedding possibly nonterminating imperative programs. In purely functional languages such as Gallina the usual approach  is to encode imperative features with monads~\cite{Moggi89}. We use a combination of the termination, state, and reader monads. The first one  is used for possibly nonterminating computations, the second one for stateful computations where the state may change, and the third one for state-aware computations that do not change the state, such as checking the condition of while loops.

A monad consists of a type constructor equipped with an operation usually called  \texttt{ret} for trivial computation, and another one,  usually called \texttt{bind}, for sequencing computations. Each particular monad also comes with specific operations.
Assume a context where a type \texttt{T}  is declared. The type constructor for the reader monad is defined as follows:
\begin{verbatim}
Definition reader (A: Type): Type := T -> A.
\end{verbatim}

\noindent where \texttt{A} is a type for values returned by a computation.
Intuitively, \texttt{reader} is a side-effect free function from states to computed values. Unlike the state monad, shown below, it only reads from states, without modifying them; hence the monad's name \texttt{reader}.

 Trivial computation  \texttt{ret} consists in ignoring the state:

\begin{verbatim}
Definition ret {A: Type}(a: A) : reader A := fun _ => a.
\end{verbatim}

\noindent Placing the parameter \texttt{A} between curly brackets marks it as \emph{implicit} - its value can therafter be automatically inferred, relieving users from the burden of having to instantiate it.

 Sequencing of computations consists in passing the result of the first computation to the second one:

\begin{verbatim}
Definition rbind {A B: Type}(m: reader A)(f: A -> reader B): reader B :=
fun s => f (m s) s.
\end{verbatim}

\noindent Finally, the reader monad is equipped with a primitive for reading the state:

\begin{verbatim}
Definition get: reader T := fun s => s.
\end{verbatim}

\noindent The other monad used in  the shallow  embedding of an imperative language in Coq is a combination of   termination and state monads, for which we use the  \texttt{option}  type constructor:

\smallskip

\begin{verbatim}
Inductive option (X: Type): Type := None: option X | Some: X -> option X.
\end{verbatim}

\smallskip

\noindent where \texttt{None} encodes nontermination. Termination state monads are the programs of our imperative language, which is reflected in the name of  its  type constructor:

\smallskip

\begin{verbatim}
Definition program (A: Type): Type := T -> option (A * T).
\end{verbatim}

\smallskip

\noindent The reader monad is a special case of the termination state monad. Thus we introduce a \emph{coercion} that  enables Coq to automatically convert a reader monad into a termination state monad when needed:

\smallskip

\begin{verbatim}
Coercion reader_to_program {A: Type}(m: reader A): program A :=
fun s => Some (m s, s).
\end{verbatim}

\smallskip

\noindent The \texttt{ret} and the \texttt{get} primitive of the reader monad are thus automatically converted. But the sequencing of computations needs to be redefined in order the take into account the fact that the first of  the sequenced computations might change the state, or might not terminate:

\smallskip

\begin{verbatim}
Definition bind {A B: Type}(m: program A)(f: A -> program B): program B :=
fun s => match m s with None => None | Some (a, s') => f a s' end.
\end{verbatim}

\smallskip

\noindent The termination state monad comes with an additional primitive to change the state:

\smallskip

\begin{verbatim}
Definition put (s: T) : program unit := fun _ => Some (tt, s).
\end{verbatim}

\smallskip

\noindent The \texttt{unit} type (inhabited by one term - the constant \texttt{tt}) is used in our functional-laaguage setting for modelling imperative programs that do not return anything, encoded by returning the dummy constant~\texttt{tt}.
 
 As a running example we  consider an imperative program that computes the length of a linked list.
First we need to specify the  \texttt{State} on which the program works. We assume a machine with two positive integer registers and an unbounded memory whose adresses and stored values are also positive integers:

\smallskip

\begin{verbatim}
Record State: Type := {reg1: nat; reg2: nat; memory: nat -> nat}.
\end{verbatim}

\smallskip

\noindent Then, using the primitives of  our monad we write  operations to read/write in registers and memory:
\begin{itemize}
\item \texttt{read_reg1},  \texttt{write_reg1} for reading, resp.\ writing in the first register, and similar functions for reading/writing in the second register. The second register is increased by one using \texttt{incr_reg2};
\item \texttt{do next <- read_addr curr} assigns to  \texttt{next}  the content of the address \texttt{curr} in memory.
More generally, reading operations return a value
that can be bound to an identifier using the standard "do" notation of monads, i.e., \texttt{do x <- m; f x} is a shortcut for
\texttt{rbind m (fun x => f x)};

\item operations are sequenced by double semicolons: \texttt{m;;f} is a notation for\, \texttt{bind m (fun _ => f)}.
\end{itemize}
This almost allows us to write a program computing the length of a linked list. Not completely, because the program uses a  \texttt{while} loop, which is not  defined until later in the paper:

\smallskip

\begin{verbatim}
Definition length (addr: nat): program State nat :=
write_reg1 addr;;
write_reg2 0;;
while (do curr <- read_reg1; ret  (curr != 0))
  (
   incr_reg2;;
   do curr <- read_reg1;
   do next <- read_addr curr;
   write_reg1 next
  );;
do res <- read_reg2; ret res.
\end{verbatim}

\smallskip

\noindent It is assumed that the linked list\footnote{For simplicity we consider linked lists where each element only contains the  next element's address.} of interest starts in memory at address
\texttt{addr}. This address is written into the first register, then the second register (which is to contain the length of the list) is initialized to
zero. Next, while the current address of the first register is not \texttt{null} (also encoded by zero), the second register is incremented and the first register is updated to point to the next element of the linked list. Finally,  at the end of the  \texttt{while} 
loop  (if the end is ever reached), the value in the second register is the length of the list of interest, hence, it is the value returned by our function whenever it terminates.

What is still missing is, of course,  the definition of the \texttt{while} loop. A first attempt uses recursion:

\smallskip

\begin{verbatim}
Fixpoint 
 while{T: Type}(cond: reader T bool)(body: program T unit): program T unit:=
  do c <- cond;if c then body;;while cond body else reader_to_program (ret tt)
\end{verbatim}

\smallskip

\noindent That is, a recursive function (introduced by the keyword  \texttt{Fixpoint}) attempts to define a \texttt{while} loop (with condition
\texttt{cond} and body \texttt{body}  of appropriate types) by first checking the condition, and if the condition holds, executing the body then 
recursively the \texttt{while} loop; otherwise, doing nothing (which is encoded by \texttt{reader_to_program (ret tt)}).  However, this function does not always terminate.
For example, if the  \texttt{while} loop  is used to navigate a linked list, like in the case of the \texttt{length} function above, and the 
list is badly linked, i.e., it contains a loop, then the   \texttt{while} loop does not terminate.
Coq rejects this  definition attempt, as it rejects any recursive function whose termination it cannot  infer.

 In the rest of the paper we show how possibly infinite while loops (and, in general, partial recursive function) can be accepted by Coq
 by encoding nontermination as evaluation to a special value.

\section{Partial Recursive Functions in Coq}
\label{sec:recursion}
 
The \texttt{while} loop is a particular case of a partial recursive function. We first sketch how partial recursive functions can be encoded in Coq before providing details.

\subsection{Outline of the Approach}
\label{sec:outline}
Assume we want to define a partial recursive function \texttt{f} from type \texttt{A} to type \texttt{B}.  A natural way to proceed is to 
give \texttt{f} the type \texttt{A -> option B}, where for any \texttt{a:A},  \texttt{f a = None} encodes the fact that \texttt{f} is undefined for
the input \texttt{a}.  
In order to define a function we further need its \emph{functional}, an abstract representation of the body of the function 
being defined.  Let \texttt{F : (A -> option B)-> A -> option B} be the functional for \texttt{f}. We say  that
\texttt{f := F f} is the \emph{fixpoint definition} of \texttt{f}.  The interesting case we here solve is when fixpoint definitions are not accepted by Coq - just like in the case of the \texttt{while} function above.

We proceed as follows. We define an auxiliary function \texttt{f_fuel :  nat -> A  -> option B} 
with an additional natural-number parameter  called the \emph{fuel}, as the following recursive function, which is accepted by Coq because Coq ``sees'' that  the \texttt{fuel} parameter  strictly decreases at each recursive call:

\begin{verbatim}
Fixpoint f_fuel (fuel: nat) (a: A) : option B:=
match fuel with
|S fuel' => F(f_fuel fuel') a
    (*S is the successor function on natural numbers*)
|0 => None
end
\end{verbatim}

\noindent If the functional \texttt{F} is  \emph{monotonic}  then, based on results in order theory explained later in this section, the  function  \texttt{f_fuel}
 can be \emph{lifted} to a \emph{continuous}  function  in \texttt{conat -> option B} where \texttt{conat}
is the type 

\begin{verbatim}
Inductive conat: Type:= finite: nat -> conat | infinity
\end{verbatim}

\noindent That is, the  inhabitants of \texttt{conat} are natural numbers wrapped with the \texttt{finite} constructor, together with the constant \texttt{infinity}. 
 Putting back the parameter \texttt{a:A} in the type we obtain a function  \texttt{f_inf} of the type \texttt{conat -> A -> option B}.  The results  later in the section also ensure that, under an additional condition on \texttt{F} (preservation of continuity),  the function  \texttt{(f_inf infinity)} is the \emph{least fixpoint} of~\texttt{F}.
 
 Recapitulating, we started with the intention of defining a function  \texttt{f : A -> option B},  using its functional \texttt{F : (A -> option B)-> A -> option B}, via the fixpoint definition \texttt{f := F f}. We have assumed this is rejected by Coq.  Per the results below
we define
 \texttt{f:= f_inf infinity}  and prove that is the least solution of  the fixpoint equation  \texttt{f = F f} - precisely the solution that Coq would have constructed had it accepted  the definition \texttt{f := F f} - with the advantage that our definition  \emph{is} accepted.

\subsection{Elements of Order Theory}
The results in this subsection have been adapted from the textbook~\cite{DBLP:books/daglib/0093287}. We have formalized them in Coq, hence, hereafter proofs 
are only sketched or omitted altogether. The examples are not only used for illustration purposes: they also serve as building blocks in our approach to partial recursive functions.

\begin{definition}
\label{def:ppo}
A \emph{pointed partial order} (PPO)  $(S, \preceq,\bot)$  is a partially ordered set $(S, \preceq)$ together with a distinguished element $\bot \in S$
such that for all $s \in S$, $\bot \preceq s$.
\end{definition}

\begin{example}
\label{ex:natppo}
{\rm 
The triple $(\N,\leq, 0)$ consisting of natural numbers $\N$, their usual order $\leq$,  and the least natural number  $0$ form a PPO. 
}
\end{example}

\begin{example}
\label{ex:flatppo}
{\rm
For any set $A$, the triple $(A \cup\{\bot\}, \preceq,\bot)$ with $\bot \notin A$, and $\preceq$ being defined as the smallest relation on $A \cup\{\bot\}$ such that  $\bot \preceq a$  and $a \preceq a$  for all $a\in A$, is a PPO called the \emph{flat PPO of A}.
}
\end{example}

\noindent In Coq the flat PPO of a type \texttt{A} is encoded using the type \texttt{option A} where \texttt{None}  plays the role of $\bot$.

\begin{definition}
\label{def:directed}
Given  a PPO $(S, \preceq,\bot)$, a set $S' \subseteq S$ is \emph{directed} if $S \not = \emptyset$ and for all $x,y \in S$ there
exists $z \in S$ such that $x, y \leq z$.
\end{definition}

\begin{example}
\label{ex:directednat}
{\rm
Any nonempty set of natural numbers in the PPO  $(\N,\leq, 0)$  is directed. 
}
\end{example}

\noindent The above example  is a consequence of the more general fact that
any nonempty sequence, i.e., totally ordered subset of elements 
in a PPO, is directed. Indeed, directed sets  are generalizations of  sequences.

\begin{example}
\label{ex:directedflat}
{\rm 
In a flat PPO $(A \cup\{\bot\}, \preceq,\bot)$, the directed sets are exactly: the singletons $\{x\}$ with $x \in A \cup\{\bot\}$,  and the pairs of elements 
of the form $\{a,\bot\}$ with $a \in A$.
}
\end{example}

\begin{definition}
\label{def:cpo}
A \emph{Complete Partial Order} (CPO) is a PPO $(S, \preceq,\bot)$ with the additional property that any directed set
$T \subseteq S$ has a \emph{least upper bound}, denoted  by $\lub\, T$.
\end{definition}

\noindent Least upper bounds of directed sets are generalizations of limits of   sequences.

\begin{example}
\label{def:conat}
{\rm
Consider the PPO  $(\N \cup \{\infty\},\leq, 0)$ with the order $\leq$ on natural numbers extended such that $\infty \leq \infty$
and $n \leq \infty$ for all $n \in \N$. Then,   $(\N \cup \{\infty\},\leq, 0)$  is a CPO. Indeed, in this totally ordered set all 
subsets $T \subseteq \N \cup \{\infty\}$ are directed, and $\lub\, T$ is either:
\begin{itemize}
\item the maximum of $T$, if it exists
\item $\infty$, if the maximum of $T$ does not exist.
\end{itemize} 
}
\end{example}

\noindent In Coq the set $\N \cup \{\infty\}$  shall be encoded as the type \texttt{conat} seen earlier in this section.

\begin{example}
{\rm
In a PPO $(A \cup\{\bot\}, \preceq,\bot)$, the least upper bound of a singleton $\{x\}$ is $x$
and the least upper bound of a pair $\{\bot, a\}$ is $a$. Those are  the only directed sets in this PPO;
hence,  $(A \cup\{\bot\}, \preceq,\bot)$ is a CPO.
}
\end{example}

\noindent Informally, the notion of compactness below captures what it means for an element to be finite.

\begin{definition}
\label{def:compact}
In a CPO $(S, \preceq,\bot)$, an element $s^\circ \in S$ is \emph{compact} whenever for all directed sets 
$T \subseteq S$, if $s^\circ \preceq \lub\, T$ then there exists $t \in T$ such that $s^\circ \preceq t$.
\end{definition}

\begin{example}
\label{ex:conatcompacts}
{\rm
 In the CPO  $(\N \cup \{\infty\},\leq, 0)$ the compact elements are exactly the (finite)  natural numbers.
}
\end{example}

\begin{example}
\label{ex:optioncompacts}
{\rm
 In the CPO  $(A \cup \{\bot\},\preceq, \bot)$ all the elements are compact.
}
\end{example}

\noindent Some CPOs are, in the following sense, completely determined by their compact elements:

\begin{definition}
\label{def:algebraic}
A CPO $(S,\preceq,\bot)$ having the set $S^\circ \subseteq S$ of compacts is \emph{algebraic} if for all $s \in S$,
the set $\{s^\circ \in S^\circ  \  \mid \ s^\circ \preceq  s \}$ is directed, 	and $s = \lub\ \{s^\circ \in S^\circ  \  \mid \ s^\circ \preceq  s \}$.
\end{definition}

\begin{example}
\label{ex:conatalgebraic}
{\rm
The CPO  $(\N \cup \{\infty\},\leq, 0)$ is algebraic. Indeed, for all $n \in \N \cup \{\infty\}$, the set of compacts (natural numbers)
$\{m \in \N \  \mid \  m \leq n \}$ is directed (as is any nonempty subset of $\N \cup \{\infty\}$). Moreover,
\begin{itemize}
\item if $n = \infty$, then the set $\{m \in \N \  \mid \  m \leq n \}$ coincides with $\N$, and $\lub \, \N = \infty$;
\item if $n \in \N$, then $\lub \, \{m \in \N \  \mid \  m \leq n \} = n$.
\end{itemize}
}
\end{example}

\begin{example}
\label{ex:flatalgebraic}
{\rm
The flat CPO  $(A \cup \{\bot\},\preceq, \bot)$ is algebraic. Indeed:
\begin{itemize}
\item  for all $a \in A$, the set $\{\bot, a\}$ of compacts in the 
$\preceq$ relation with $a$ is directed, and $a = \lub \, \{\bot, a\}$
\item the set $\{\bot\}$ of compacts in the 
$\preceq$ relation with $\bot$ is directed and $\lub \, \{\bot\}  = \bot$.
\end{itemize}
}
\end{example}

\noindent We shall use the following notion, which relates a PPO to the compact elements of an algebraic CPO:

\begin{definition}
\label{def:embedding} Consider a PPO $(S^{\!\circ},\preceq^\circ,\bot^{\!\!\circ})$ and an algebraic CPO $(T, \preceq,\bot)$ whose set 
of compacts is $T^\circ$. We say that $(T, \preceq,\bot)$ is an \emph{embedding} of $(S^{\!\circ},\preceq^\circ,\bot^{\!\!\circ})$ if there exists
an injection  $\iota : S^{\!\circ} \to T$ such that:
\begin{itemize}
\item $\iota \, \bot^{\!  \!\circ} = \bot$;
\item $\iota$ is monotonic;
\item 
$\iota$ maps $S^\circ$ to $T^\circ$, written  $\iota \, S^{\!\circ} = T^\circ$.
 \end{itemize}
 The embedding, including the  injection involved in it, is denoted by  
 $\iota : (S^{\!\circ},\preceq^\circ,\bot^{\!\!\circ}) \to  (T, \preceq,\bot)$.
\end{definition}

\begin{example}
\label{ex:embeddings}
{\rm
The embedding $\kappa : (\N,\leq, 0) \to (\N \cup \{\infty\},\leq, 0)$ is induced by the canonical inclusion $\kappa$ of $\N$ into $\N\cup  \{\infty\}$. The embedding of $(A \cup \{\bot\}, \preceq, \bot)$ into itself is also induced by the canonical inclusion.
}
\end{example}

\noindent Remark:  not all embeddings are induced by canonical inclusions. In Coq, in general, they are  not. This is because Coq is based on
type theory, hence, one cannot just add a new element to a type; to do this one must create a new type and wrap the old type in a constructor (which, in Coq, is always an injection). An example of this is the representation of $\N \cup \{\infty\}$ as the type
\texttt{conat} with a constructor \texttt{finite : nat -> conat}. With the appropriate order and  bottom element, \texttt{conat}  is an embedding of \texttt{nat} induced by the  constructor \texttt{finite}.
The only situations when canonical inclusion induces an embedding in Coq  is
when it coincides with the  identity function, like in the  embedding of $(A \cup \{\bot\}, \preceq, \bot)$ into itself.

\begin{definition}
\label{def:revinj}
Assume an embedding $\iota : (S^{\!\circ},\preceq^\circ,\bot^{\!\!\circ}) \to  (T, \preceq,\bot)$.
We denote by $\iota^{-1} : T \to S^{\!\circ}$ the (unique) function such that $\iota^{-1} \,  (\iota\,  s^\circ) = s^\circ$ for all
$s^\circ \in S^{\!\circ}$,
and $\iota^{-1} \, t = \bot^{\!\!\circ}$ for $t \in T \setminus (\iota \, S^{\!\circ})$.
\end{definition}

\noindent Hence $\iota^{-1}$ is the inverse of $\iota$ on the compacts $\iota \, S^{\!\circ}$ of $T$, and elsewhere it is given the (arbitrary) 
value $\bot$.

The next theorem is our main  ingredient for defining and reasoning about partial recursive functions. It uses the following notion of continuity:
\begin{definition}
\label{def:cont}
Given two CPOs $(T, \preceq,\bot)$ and  $(T', \preceq' ,\bot')$, a function
$f : T \to T'$ is \emph{continuous} if for any directed set $S \subseteq T$, its image $(f \ S) \subseteq T'$
is directed, and $f \ (\lub \ S) = \lub(f \ S)$.
\end{definition}

\begin{theorem}
\label{th:lifting}
Assume two embeddings $\iota_1 : (S^{\!\circ}_1,\preceq^{\circ}_1,\bot^{\!\!\circ}_1) \to  (T_1, \preceq_1,\bot_1)$
and  $\iota_2 : (S^{\!\circ}_2,\preceq^{\circ}_2,\bot^{\!\!\circ}_2) \to  (T_2, \preceq_2,\bot_2)$ and a monotonic function
$f^\circ : S^{\!\circ}_1 \to S^{\!\circ}_2$. Then there exists a unique continuous function 
$f : T_1 \to T_2$ such that $f = \iota_ 2 \, \circ f^\circ \,  \circ \, \iota^{-1} _1$   \ --- where $\circ$ is the standard notation for function composition.
\end{theorem}

\noindent If the embeddings in Theorem~\ref{th:lifting} are canonical inclusions we have a simpler version of the above result:

\begin{corollary}
\label{cor:lifting}
Assume two embeddings $\iota_1 : (S^{\!\circ}_1,\preceq^{\circ}_1,\bot^{\!\!\circ}_1) \to  (T_1, \preceq_1,\bot_1)$
and  $\iota_2 : (S^{\!\circ}_2,\preceq^{\circ}_2,\bot^{\!\!\circ}_2) \to  (T_2, \preceq_2,\bot_2)$ where $\iota_1, \iota_2$ are canonical inclusions.
Then, for any any monotonic function
$f^\circ : S^{\!\circ}_1 \to S^{\!\circ}_2$ there exists a  unique  continuous function 
$f : T_1 \to T_2$ such that for all $s^\circ \in S^{\!\circ}_1$, $f \ s^\circ  = f^\circ s^\circ$.
\end{corollary}

\subsection{Application to Partial Recursive Functions}
\label{sec:appl}
We use the existence part of Theorem~\ref{cor:lifting} in order to define partial recursive functions and the uniqueness part in order to
prove that the defined functions are least fixpoints of their respective fixpoint equations. 

The method has been formalized in Coq; we sketch it below in mathematical notation. Assume that we want to define a partial function $f : A \to  (B \cup \{\bot\})$. We have have at our disposal the 
functional $F : (A \to (B \cup \{\bot\})) \to A \to (B \cup \{\bot\})$.  The following assumptions on $F$ are required:

\begin{itemize}
\item monotonicity: for all $f, f' :A \to  (B \cup \{\bot\})$,  if  \emph{for all $a \in A$, $f \, a \preceq f' a$} then
\emph{for all $a \in A$, $F \, f \, a \preceq F \, f'  a$}, where $\preceq$ is the flat order on $(B \cup \{\bot\})$. 

\item preservation of continuity: assume an arbitary function $g : (\N \cup \{ \infty\}) \to A \to  (B \cup \{\bot\})$. If, for each 
$a \in A$, the function $\lambda \, n \to g \ n\ a$ is continuous (as a function  between 
$(\N \cup \{ \infty\})$  and  $(B \cup \{\bot\})$ organized as CPOs) then, for each $a' \in A$ the function 
 $\lambda\, n \to F \ (g \ n) \ a'$ is continuous as well.
\end{itemize}

 The method proceeds as a series of steps, grounded  in the results from the previous subsection:

\begin{enumerate}
\item A function $f^\circ : \N \to A \to (B \cup {\bot})$ is recursively defined by the equations: for all $a \in A$ and $m \in \N$, $f^\circ \ 0 \ a = \bot$ and 
 $f^\circ \ (m +1)  \  a =  F \ (f^\circ \  m) \ a$; intuitively,  for all $m \in \N$, $(f^\circ \, m)$  constitute approximations of the function that we want to define, constrained by the finite amount of  fuel $m \in \N$;
 
 \item the \emph{monotonicity} requirement on $F$ ensures that, for all $a \in A$, the function 
 $f^{\circ}_a = \lambda \, n \to (f^\circ  \, n \ a)$ is monotonic as a function between 
$\N$  and  $(B \cup \{\bot\})$ organized as PPOs;

 \item the \emph{existence} result of Theorem~\ref{th:lifting} ensures that, for all $a\in A$ that there exists a continuous function
 $f_a:  (\N \cup \{ \infty\}) \to (B \cup \{\bot\})$, satisfying $f_a \ m = f^\circ \, m  \ a$ for all $m \in \N$;
 here, we have used the fact that, in the embeddings involved in our application of Theorem~\ref{th:lifting}, the injections are the canonical inclusions
 (cf.\  Example~\ref{ex:embeddings}), hence,  one can apply the simpler version of the theorem --- Corollary~\ref{cor:lifting}; 
 
 \item using the \emph{uniqueness} result of the corollary, for all $a \in A$, any 
 continuous function  $f'_a:  (\N \cup \{ \infty\}) \to (B \cup \{\bot\})$ satisfying
  $f'_a \ m  = f^\circ \, (m + 1)  \ a$ for all $m \in \N$,
 also satisfies $f'_a = f_a  \circ (\lambda\,  n \to n + 1)$; here we have used the fact that 
 the function  $\lambda \ n \to n + 1 :  (\N \cup \{ \infty\}) \to  (\N \cup \{ \infty\})$ is continuous, and 
 that the composition of continuous functions is a continuous function as well;
 
 \item for all $a \in A$,  let  $f'_a:  (\N \cup \{ \infty\}) \to (B \cup \{\bot\})$ be defined by
  $f'_a = \lambda \,  n  \to   F \ (\lambda\, (x:A) \to f_x \  n) \ a$.  Using the \emph{continuity preservation} requirement on $F$,
  the continuity of $f'_a$ reduces to the continuity of $\lambda \, n \to (\lambda\, (x:A) \to f_x \  n)  \ a  = \lambda \, n \,\to f_a \ n = f_a$; since $f_a$ is continuous, $f'_a$ is continuous as well;
  
  \item moreover, using the definitions of $f^\circ$ (first item in this list), of $f_a$ (item 2),  and of $f'_a$ (item 5):  for all $m \in \N$, $f'_a \ m = F \ (\lambda x \to f_x \  m) \   a = F ( \lambda \, x \to f^\circ \, m  \ x) \ a = 
   F \  (f^\circ \, m ) \ a = f^\circ \ (m +1)  \  a$; hence, (cf. item~4), $f'_a = f_a  \circ (\lambda\,  n \to n + 1)$. This implies that for all $n \in (\N \cup \{ \infty\}) $, $f_a \ (n + 1) = F \ (\lambda x \to f_x \  n)$\, ;
 
 \item let now $f : A \to  (B \cup \{\bot\})$ defined, for all $a \in A$,  by $f \ a = f_a \   \infty$. Then, using item 6 and $\infty = \infty +1$: for all $a \in A$,
  $f \ a = f_a \   \infty  = f_a \   (\infty + 1) =  F \ (\lambda\,  x \to f_x \  \infty)\  a = F \ (\lambda\,  x \to f \, x)\ a = F \ f \ a$; that is, 
  we have obtained the fixpoint  equation $f = F \ f$. What remains to be proved is that
  $f$ is its least solution;
  
  \item for this, we inductively define a sequence of functions in $A \to  (B \cup \{\bot\})$ by 
  $F^0 = \lambda\, x \to \bot$ and, for all $m \in \N$,  $F^{m+1} = F (F^{m})$. Using the definition of $f$
  we prove the equality $f = \lub\, \{F^n \ | \ n \in \N \}$, where the least upper bound is taken in the CPO of
  functions $A\to (B \cup \{\bot\})$ ordered pointwise. Finally, we use a result that says that
  if $F$ is monotonic on a CPO (it is, in our case, by the \emph{monotonicity} assumption) and $ \lub\, \{F^n \ | \ n \in \N \}$
  is a fixpoint of $F$ (it is, in our case, since  $\lub\, \{F^n \ | \ n \in \N \} = f$ and $f = F \ f$) then 
  $ \lub\, \{F^n \ | \ n \in \N \} $ is the least fixpoint of $F$; i.e., $f$ is the least fixpoint of $F$. 
 \end{enumerate}
 A comparison between the proposed approach for defining  partial recursive functions and the standard one based on Kleene's fixpoint theorem is discussed in Section~\ref{sec:rel} dedicated to related works.

\subsection{Instantiation  to While Loops}
The results from the previous subsection are now instantiated to \texttt{while} loops. 

Recall from subsection~\ref{sec:outline} the failed attempt at defining \texttt{while} loops in Coq, and notice their type:

\smallskip

\begin{verbatim}
Fixpoint while{T:Type}(cond:reader T bool)(body:program T unit):program T unit:=
  do c <- cond; if c then body ;; while cond body else reader_to_program (ret tt)
\end{verbatim}

\smallskip

\noindent In order to instantiate the method described in the previous subsection to \texttt{while} loops we first need to 
change their type to \texttt{A->option B} for appropriate \texttt{A}, \texttt{B}.
Remembering that \texttt{program T unit} is defined as 
\texttt{T->option(unit*T)}, once the implicit parameter  \texttt{T}  is chosen, the type of   \texttt{while}   becomes

\smallskip

\begin{verbatim}
(reader T bool) -> (program T unit) -> T -> option(unit*T)
\end{verbatim}

\smallskip

\noindent In order to obtain a type of the form \texttt{A->option B} we \emph{uncurry} the above type to

\smallskip

\begin{verbatim}
((reader T bool)*(program T unit)*T) -> option(unit*T)
\end{verbatim}

\smallskip

\noindent where \texttt{*} builds products between types. 
 Next, we define a function \texttt{while':A->option B} where 
\texttt{A=(reader T bool)*(program T unit)*T} and  \texttt{B=unit*T}.  For this we first write the  \emph{functional} for the \texttt{while'} function 
as follows

\smallskip

\begin{verbatim}
Definition While'{T:Type}(W:((reader T bool)*(program T unit)*T)->option unit*T)
                        (p:(reader T bool)*(program T unit)*T): option unit*T :=
let (cond,body,s)  := decompose p in                
(
 do c <- cond;
 if c then 
 body;;(fun (s':T)=>W (cond, body, s')) (*after ;; a function on T is expected*)
 else reader_to_program (ret tt)
) s
\end{verbatim}

\smallskip

\noindent (Notice how the parameter \texttt{p} was decomposed into three components.) After proving that \texttt{While'} is monotonic and preserves continuity, we obtain using the method in Subsection~\ref{sec:appl}
the function \texttt{while': (reader T bool)*(program T unit)*T -> option unit*T} as the least fixpoint of \texttt{While'}.  

What remains to be done is to \emph{curry}  the type \texttt{(reader T bool)*(program T unit)*T->option unit*T} to the expected 
 type of the \texttt{while} function. When this is done, we obtain  \texttt{while\{T:Type\}: (reader T bool)->(program T unit)->T->option unit*T} as the least fixpoint of the functional
 
 \medskip
 
\begin{verbatim}
Definition While{T:Type}(W:(reader T bool)->(program T unit)->T->option unit*T)
     (cond :(reader T bool))(body : (program T unit))(s :T)) : option unit*T :=
(do c <- cond; if c then body;;(W cond body) else reader_to_program (ret tt)) s
\end{verbatim}

\medskip

\noindent which concludes our construction of \texttt{while} loops in Coq. The next step is to provide users with means to reason about programs that contain such loops. This is the object of the next two sections. They shall be using the two following facts, which are consequences of \texttt{while} being the least fixpoint of its functional:

\begin{itemize}
\item an unfolding lemma, which is just another form of the fixpoint equation:

\begin{verbatim}
Lemma while_unfold{T:Type}: forall(c: reader T bool)(b: program T unit),
while c b =
      (do c' <- c; if c' then b;;while c b else reader_to_program (ret tt))
\end{verbatim}
        
\item  a lemma stating that the \texttt{while} loop evaluates to
\texttt{Some x} in a state if and only if there exists a fuel-constrained version of 
the loop that also evaluates to the same \texttt{Some x} in the same state:

\begin{verbatim}
Lemma while_iff_while_fuel{T:Type}:
forall (c:reader T bool)(b:program T unit)(s:T)(x:unit*T), 
while c b s = Some x <-> exists (fuel:nat), while_fuel fuel c b s = Some x.
\end{verbatim}

Since evaluation to \texttt{Some x} models termination, the  lemma can also be read as``if a loop terminates, then it terminates in
finitely many steps'' - where the number of steps is upper-bounded by  \texttt{fuel}.
\end{itemize}

\section{Partial Correctness}
\label{sec:correctness}

In this section we define a monadic Hoare logic for partial correctness~\cite{Hoare69,Swierstra09}.
Roughly speaking, partial correctness expresses the fact that a program returns the right answer whenever it terminates. In this paper, a program is a monadic computation.
Remembering that \texttt{Prop} is the Coq type for logical statements,  one writes the Hoare triple \texttt{\{\{P\}\}\,m\,\{\{Q\}\}}  for the proposition \texttt{hoare\_triple\;P\;m\;Q} defined in Coq by
\smallskip

\begin{verbatim}
Definition hoare_triple
  {T A : Type} (P : T -> Prop) (m : program T A) (Q : A -> T -> Prop) : Prop :=
forall s s' a, P s -> m s = Some (a, s') -> Q a s'.
\end{verbatim}

\smallskip
\noindent 
That is, if the
program \texttt{m:\,(program T A)} is in a state \texttt{s:\,T} such that the pre-condition \texttt{P: T -> Prop} holds for \texttt{s}, if the program terminates (encoded by the fact that the program returns \texttt{Some\,(a, s')}) then the pair \texttt{(a, s')} satisfies the postcondition \texttt{Q: A -> T -> Prop}.

There are Hoare triples for all monadic instructions, but the triple of interest is this paper is the one for the \texttt{while} loops. It states that: if the  \texttt{body} of the loop preserves an invariant \texttt{I} as long as the condition \texttt{cond} of the loop is true, then the loop preserves the invariant whenever it terminates.

\smallskip

\begin{verbatim}
Lemma while_triple
  {T: Type}(cond: reader T bool)(body: program T unit)(I: T -> Prop):
{{ fun s => I s /\ cond s = true }} body {{ fun _ s' => I s' }} ->
{{ I }} while cond  body {{ fun _ s' => cond s' = false /\ I s' }}.
\end{verbatim}

\smallskip

\noindent In order  to prove  \texttt{while_triple} we first prove a triple for fuel-constrained loops by induction on
\texttt{fuel}:

\smallskip

\begin{verbatim}
Lemma while_fuel_triple
 {T:Type}(fuel:nat)(cond: reader T bool)(body: program T unit)(I: T -> Prop):
{{ fun s => I s /\ cond s = true }}body{{ fun _ s' => I s' }} ->
{{ I }} while_fuel fuel cond body  {{ fun _ s' => cond s' = false /\ I s' }}.
\end{verbatim}

\smallskip

\noindent The lemma  \texttt{while_fuel_triple} is used in the proof of \texttt{while_triple}, together with the lemmas
 \texttt{while_iff_while_fuel} and \texttt{while_unfold} shown at the end of the previous subsection.
 
Then,  \texttt{while_triple} is used to prove the \emph{weakest precondition} triple  for our running example:

\smallskip

\begin{verbatim}
Lemma length_wp (addr: nat)(P: nat -> State -> Prop):
{{ fun s => forall len,
     Length s addr len ->
     P len {| reg1 := 0; reg2 := len; memory := s.(memory) |} }}
length addr
{{ P }}.
\end{verbatim}

\smallskip

\noindent where \texttt{Length} is an inductively defined relation that says that in a state \texttt{s}, the  list starting at address \texttt{addr} has length \texttt{len}. Having this predicate gives us an abstract manner of defining a linked list's length:

\smallskip

\begin{verbatim}
Inductive Length (s:State): nat -> nat -> Prop :=
| length_nil : forall addr, addr = 0 -> Length s addr 0
| length_cons :
  forall addr len,
  addr <> 0 -> Length s (s.(memory) addr) len -> Length s addr (S len).
\end{verbatim}

\smallskip

\noindent In accordance to the laws of weakest precondition,  \texttt{length\_wp}  says that the postcondition \texttt{P} must hold in the precondition when applied to
  the expected return value upon termination (the length \texttt{len}) and to  the expected state upon termination \texttt{\{|reg1 := 0; reg2 := len; memory := s.(memory)|\}}.
%
%This allows us to transform any postcondition \texttt{P : nat -> State -> Prop} into a precondition.
%

As usual with Hoare logic, the crux of the proof is to find the right loop invariant to be fed to the lemma \texttt{while_triple}. Here, it is a generalization of the precondition in  \texttt{length\_wp}:

\smallskip

\begin{verbatim}
fun _ s => forall len,
  Length s s.(reg1) len ->
  P (len+s.(reg2)) {| reg1:= 0; reg2:= len + s.(reg2) ; memory:= s.(memory) |}).
\end{verbatim}

\smallskip

\noindent 
With this choice of invariant the proof of \texttt{length\_wp} is just a matter of unfolding definitions.

The interest of having proved a weakest precondition lies in its generality. As immediate corollaries of \texttt{length_wp} we obtain a first lemma that states that whenever \texttt{length addr} terminates, the register \texttt{reg2} contains the length of the linked list starting at address \texttt{addr}:

\smallskip

\begin{verbatim}
Lemma length_correct1 (s0: State)(addr: nat) :
{{ fun s => s = s0 }} length addr {{ fun _ s' => Length s0 addr s'.(reg2) }}.
\end{verbatim}

\smallskip

\noindent And another lemma stating that: if the linked list starting at address \texttt{addr} has length \texttt{len}, then this is the value that will be returned by \texttt{length addr} whenever it terminates.

\smallskip

\begin{verbatim}
Lemma length_correct2 (len: nat)(addr: nat) :
{{ fun s => Length s addr len }} length addr {{ fun n _ => n = len }}.
\end{verbatim}

\section{Termination}
\label{sec:termination}

In our approach, termination is modeled by evaluation to \texttt{Some} value. In order to succesfully prove termination we need to
specify quite precisely
the value to which a program evaluates. The key for the termination of the \texttt{length} program is the termination of its \texttt{while}
loop, which is expressed as follows:

\smallskip 
\begin{verbatim}
Lemma while_terminates:
forall len s addr, Length s addr len ->
forall n, 
while do curr <- read_reg1; ret  (curr != 0)
(incr_reg2;; do curr <- read_reg1; do next <- read_addr curr; write_reg1 next)
{| reg1 := addr; reg2 := n; memory := s.(memory) |} =
Some (tt, {| reg1 := 0; reg2 := n + len; memory := s.(memory) |}).
\end{verbatim}

\smallskip

\noindent That is, when called in a state 
where the first register \texttt{reg1} points to the beginning of the list and the second register \texttt{reg2}  is initialized with some value \texttt{n}, the 
\texttt{while} loop ends in a state where the first register is \texttt{null} and the second register contains \texttt{n} plus the length of the list. The \texttt{memory} field of the state remains unchanged because the loop does not write in it.
The return value  \texttt{tt} is the unique inhabitant of \texttt{unit} and encodes the fact  that \texttt{while} loops do not actually return anything relevant.

Lemma  \texttt{while_terminates} is proved by induction on \texttt{n} and uses the lemma \texttt{while_unfold} to unfold the loop in the inductive step.
It is then used to prove the termination of the \texttt{length} program:

\smallskip

\begin{verbatim}
Lemma length_terminates: 
forall  s len addr,
Length s addr len -> 
length addr s = 
 Some (len,{|reg1 := 0; reg2 := len; memory := s.(memory)|}).
\end{verbatim}

\smallskip

\noindent This says that the function call \texttt{length addr s} terminates with value \texttt{len}  whenever, according to the inductive relation  \texttt{Length}, the state \texttt{s}  has a \texttt{memory} field where there  is a well-formed linked list of length \texttt{len} starting at address \texttt{addr}. 
If the list were not well-formed, i.e. its links would form a loop, then the function would evaluate to \texttt{None}, which, in our method denotes non-termination. 

\section{Related Work}
\label{sec:rel}
In domain theory~\cite{winskel93} partial recursive functions are typically defined as least fixpoints of their functionals
using \emph{Kleene's fixpoint theorem}, which states that a functional has a least fixpoint whenever it is \emph{continuous}, and that the least fixpoint is obtained by infinitely many iterations of the functional starting from a function defined nowhere. This theorem is very elegant and easy to prove. Our own version of a
fixpoint theorem (perhaps less elegant than Kleene's theorem, and definitely harder to prove) has the same conclusion  but requires that the functional be monotonic and \emph{continuity-preserving}, which roughly means that, given a certain continuous function, the functional produces a continuous function. 
From our own experience  and that of other authors (e.g.,~\cite{DBLP:conf/ppdp/BertotK08}) it appears that \emph{using} Kleene's theorem, which requires proving
continuity, is difficult  in practice, even in the simplest cases. For example, it took us hundreds of lines of Coq code just to prove the continuity of the successor function on natural numbers extended with infinity. By contrast, using our version of the theorem requires a proof of preservation of continuity, which  appear to be more manageable - for while-loops in a shallow embedding of an imperative programming language in Coq the proof of continuity-preservation is remarkably simple: ten lines of Coq code.  One possible reason for why continuity-preservation is easier to prove for functionals than continuity is that that the latter   refers to  the \emph{higher-order} functional itself, whereas
 the former concerns the argument of the functional, which is a simpler function, one order below the order of the functional. 
 
Other authors have explored partial recursive functions in Coq. In~\cite{capretta2005} a partial recursive function's codomain is a \emph{thunk} - a parameterized coinductive type that "promises" an answer as a value of its parameter, but may postpone this answer forever, yielding nontermination. However, the  functions being defined now become \emph{corecursive} functions, which are   restricted in the Coq proof assistant. As a result, only \emph{tail-recursive} functions can  be defined with this approach. This was also noted in~\cite[Chapter 7.3]{cpdt}. 

The same author \cite[Chapter 7.2]{cpdt} proposes an alternative for the codomain of a partial function: a \emph{computation} is a type that associates to a natural-number approximation level an approximation of the intended function.  Like us the author uses  the Coq \texttt{option}
 type, where \texttt{None} stands for nontermination and \texttt{Some} for termination with a return value; and a monad of computations is designed so that computations can
  encode imperative features. However,  
 \cite[Chapter 7.2]{cpdt} requires that the functional of the function being defined be continuous, 
 for a definition of continuity equivalent to that employed in domain theory; hence this approach is subject to the general difficulty of proving continuity of functionals.
  
Regarding the embedding of imperative languages in proof assistants for the purpose of program verification, an alternative to the shallow embedding that we here use is \emph{deep} embedding, which consist in defining the syntax, operational semantics~\cite{KLW14,Leroy09},   and program logic for the guest language in the logic of the host.  Among the many projects we here cite the Iris project~\cite{iris}, a rich environment built around  a concurrent programming language and the corresponding program logic - concurrent separation logic. 

\section{Conclusion and Future Work}
\label{sec:conclusion}
Recursive functions in Coq need to terminate for the underlying logic to be sound. Coq typically ensures 
termination  via an automatically-checkable structural decreasing
 of terms to which recursive calls apply. In more complex cases Coq can be helped by the user with termination proofs.  The termination proof  becomes a part of a function definition; a function is not defined until the termination proof is completed.
 
For various reasons falling under the general notion of separation of concerns it is
desirable to separate function definitions from  termination proofs. It is also  useful to have functions that do not terminate on some inputs. This paper proposes a new approach that achieves these desirable features.  Non-termination is  simulated as evaluation to a special value interpreted as "undefined". Under mild conditions on the function's body encoded as a higher-order functional, a possibly non-terminating function is defined and proved to be the least fixpoint of its functional according to a certain definition order. 
 
We instantiate the general approach to while-loops in an imperative language shallowly embedded in Coq. 
The shallow embedding is based on a combination of monads.
The least-fixpoint property of  the resulting while loops is a key property enabling termination and partial-correctness proofs  on imperative programs containing them. The  practicality of the approach is illustrated by proving partial correctness and termination properties on a program computing the length of linked lists. Partial correctness is expressed in Hoare logic and is proved in the standard manner, by having users 
provide a strong-enough invariant; and termination is proved by having users provide an upper bound for the number of iterations.

\paragraph*{Future Work} A promising line of future work is to extend our approach
to defining partial  \emph{co}recursive function in Coq. The idea is that codomains of such functions would be encodings of coinductive types organized as algebraic CPOs, generalizing the option types that we here used for recursive functions. Initial experiments with defining some difficult corecursive functions that go beyond Coq's builtin corecursion mechanisms  (a filter function on streams, a mirror function on Rose trees) are promising.  

The function-definition mechanism presented in this paper critically depends on functionals preserving continuity.
 A deeper understanding of the relationship between continuity-preservation and continuity, and of the perimeter where the continuity-preservation property holds, is also left for future work.

A more practical future work direction is to apply the instance of our approach to imperative-program definition and verification. Our intention is to build on our experience with proving low-level programs manipulating 
linked lists~\cite{jomaa:hal-01816830}. An interesting logic to consider in this setting is separation logic~\cite{DBLP:conf/vstte/Reynolds05}, perhaps by drawing inspiration from the shallow embedding of separation logic in Coq presented in~\cite{Sergey:PnP}. 

\nocite{*}
\bibliographystyle{eptcs}
\bibliography{biblio}

\end{document}